\documentclass{aa}
\usepackage{epsf}
\usepackage{graphicx}

\newcommand{\etal}{et~al.\ }

\def\ntd#1{\vtop{\footnotesize\hsize=\textwidth\leavevmode#1\hspace*{\fill}}}

\begin{document}

\title{Planetary Nebulae Near the Galactic Center: Identifications
\thanks{Observations made with the Burrell Schmidt of the Warner 
and Swasey Observatory, Case Western Reserve University. }
}
\subtitle{}
\author{George H. Jacoby\inst{1}
\thanks{Visiting Astronomer, Kitt Peak National Observatory,
     National Optical Astronomy Observatory, which is operated by the
     Association of Universities for Research in Astronomy, Inc., under
     cooperative agreement with the National Science Foundation.}
\thanks{Visiting Astronomer, Cerro Tololo Interamerican Observatory,
     National Optical Astronomy Observatory, which is operated by the
     Association of Universities for Research in Astronomy, Inc., under
     cooperative agreement with the National Science Foundation.} 
\and Griet Van de Steene\inst{2} }

\institute {WIYN Observatory, P.O. Box 26732, Tucson, AZ, 85726 
\and Royal Observatory of Belgium, Ringlaan 3, 1180 Brussels, Belgium
}

\offprints{G. H. Jacoby, \email{gjacoby@wiyn.org}}

\authorrunning{Jacoby \& Van de Steene}
\titlerunning{Planetary Nebulae Near the Galactic Center: Identifications}

\date{Received / Accepted}

\abstract
{We surveyed the central 4 x 4 degrees of the Galactic center for
planetary nebulae in the light of [S~III] $\lambda9532$ and found 94
PNe that were not previously known, plus 3 that were previously
identified as possible candidates. 
For 63 of these 97 objects, we obtained spectra that are
consistent with highly reddened PN while the other 34 could not be recovered
spectroscopically and remain unverified. Of the 94 candidates,
54 and 57 were detected via radio
at 3 and 6 cm, respectively. An additional 20 PNe candidates were 
found during follow-up H$\alpha$ imaging but have not yet been verified
spectroscopically. Based on the properties of IRAS sources in this
region of the Galaxy, and on the total luminosity of the Galactic bulge, 
the expected number of PNe is $\sim250$, only 50\% more
than the 160 PNe candidates now known. 
Thus, surveys for PNe in the bulge are approximately two-thirds 
complete with the remainder likely hidden behind dust.

 }
 
\maketitle

\section{Introduction and Motivation}

Estimates of the number of planetary nebulae (PNe) in the Galaxy have
always been subject to large uncertainties, ranging, for example, from
6,000 to 80,000 (Peimbert 1993). The principal obstacle in
deriving an accurate count is the very high level of extinction close to
the galactic plane, especially when looking toward the Galactic center
where a large fraction of the population is expected. Interest in the
population count stems from studies of the chemical enrichment rates
(e.g., Peimbert 1987) and comparisons between the populations
in our Galaxy and those in external systems (e.g., Jacoby 1980, Peimbert 1993).

For example, when looking at M31, the PN system is strongly concentrated
toward the nucleus (Ciardullo \etal 1989). That is, the PN density 
follows the radial
increase of starlight inward. Yet, the Galactic distribution fails to
rise accordingly, in contrast to the rapidly increasing population of
IRAS sources and OH/IR stars that have IR colors of PNe (see Figure 2
of Pottasch \etal 1988). If, in fact, the PNe were to follow the same
distribution as these sources, then a relative lifetime argument
implies that there are $\sim320$ PNe within 2 degrees of the Galactic
center (see Figure 3 of Pottasch \etal 1988).  Yet, prior to this survey, 
initiated in 1994, only 34 PNe, or 10\% of the expected number,
were known in this region.

The expectation of 320 PNe, though, is highly
approximate because not all color-selected IRAS candidates are true PNe.
Van de Steene \& Pottasch (1995) found that roughly 25\% of their IRAS
sources could be recovered in a radio survey, implying a minimum estimate
of 80 PNe for the central 2 degrees.
On the other hand, many PNe are not detectable by IRAS and so, IRAS counts
underestimate the true number of PNe. Unfortunately,
there is no reliable estimate
for this factor because neither the IRAS survey nor PN surveys are complete.
We return to this question in Section 3 where we derive another estimate
for the number of PNe in the Galactic Bulge.

Beyond the simple counting statistics of PNe, the greater value of
their study in and near the Galactic center is to use them as probes of
the kinematic and chemical history of the bulge population of stars. If
it were possible to identify many hundreds of PNe in the bulge region,
one could map out the matter distribution, including any dark matter
contributions, as was done in the galaxy NGC 5128 by Hui \etal (1995).

Perhaps most importantly, follow-up spectroscopy of PNe can tell us
the rate at which the alpha elements were enhanced near the Galactic
center. In particular, the elements O, Ne, S, and Ar survive the stellar
evolution and nucleosynthesis process unaffected from when the star was
formed (Forestini \& Charbonnel 1997). By analyzing a PN spectrum, one
can deduce the time of formation of the progenitor star, as well as its
initial chemical composition. By analyzing many PNe, a timeline for elemental
enhancements can be constructed, as was done for the LMC by Dopita
\etal (1997).  Once the chemical compositions are known for many PNe,
they can complement the stellar compositions (McWilliam \& Rich 1994)
by providing information on different elements; or, PNe can be used in
place of the more observationally challenging stellar composition 
analysis.

For these reasons, the interest in Galactic center PNe has grown since
we began our survey. The most productive general survey for Galactic
PNe has been described by Parker \& Phillipps (1998) and Parker \etal (2003) 
netting 1214 newly identified PNe in H$\alpha$.
Searches for PNe specifically toward the Galactic center 
and bulge have
also been carried out by G\'omez, Rodriguez, \& Mirabel (1997),
Beaulieu, Dopita, \& Freeman (1998), Kohoutek (2002), and Boumis \etal (2003).

In this paper, we report on the identification of many additional PNe
within a few degrees of the Galactic center that were not previously
known, thanks to a wide-field near-IR survey initially motivated by
discussions with Stuart Pottasch. Van de Steene \& Jacoby (2001) have
already reported the results of synthesis radio observations at 3 and 6 cm
that were carried out at the Australian Telescope Compact Array to improve
the determination of extinction measures for these highly obscured PNe.
In a subsequent paper, we will report the results of our spectroscopic
follow-up survey that provides the validation for most of these objects as PNe,
as well as radial velocities for kinematic studies, and in a few cases, chemical
composition estimates and ages for the PNe and their progenitors to
explore the chemical enrichment history of the Galactic bulge.

\section{Observational Approach}

The success of this survey was made possible by the novel use of
narrow-band imaging in the near-IR [SIII] line at 9532 \AA. The [OIII]
line at 5007 \AA\  is more commonly used for surveys because of its extreme 
intrinsic brightness.
The [SIII] line, though, has the attactive property of being the apparently
brightest line in
the spectra of typical PNe when the V-band extinction is between 4 and
12 mags. Beyond 12 mags of extinction, Br$\gamma$ becomes the survey line
of choice from the ground. 

As will be shown in a later paper where we
present spectra of the new PNe, the selection of the [SIII] line was
nearly optimal. We will show that using the [OIII] $\lambda 5007$ line
would have been disasterous due to the effects of reddening.
It is much less obvious, however, that H$\alpha$ would have been a poorer 
choice. In fact, as we show below, H$\alpha$ offers some advantages.
As a general rule, though, H$\alpha$ surveys for PNe suffer badly
from confusion with emission-line stars and HII regions and 
cannot be used reliably
unless another filter centered on a high excitation ion
(e.g., [SIII]) is used in conjunction.

We carried out the survey of the central 4x4 degrees using the Case
Western Reserve University (CWRU) Burrell Schmidt telescope on Kitt Peak
in August 1994 and July 1995. At that time this 0.6-m telescope had a
uniquely wide 69'x69' field of view with a 2Kx2K STIS CCD. The CCD's
21$\mu$m pixels subtend 2.028" on the sky, which is relatively coarse
sampling that limits the accuracy of any photometry or the identification
of stellar sources. Many of the Galactic center PNe, though, are slightly
extended (1" at the Galactic center corresponds to a radius of 0.019 pc), 
and so,
this deficiency was not serious.  In addition to its wide field, this
CCD was unthinned (i.e.,
it is a ``thick'' CCD), and this characteristic proved highly advantageous 
for working in the
near-IR. With the usually more desirable thinned CCDs, fringing at the
[SIII] wavelength is extremely severe and flat-fielding the images is
impossible.  Thus, despite the small size of the telescope and
the observational challenge of working from a northern site, the problem
of finding PNe in the Galactic bulge actually was extremely well-suited
to the equipment at hand.

For the on-band [SIII] images, a special filter was purchased having a
central wavelength of 9535 \AA\  and bandpass (FWHM) of 29 \AA\  in the
f/3.5 beam of the CWRU Schmidt. We used a broad RG-850 to serve as an
``off-band'' filter.

\subsection{The [SIII] Survey}

The coordinates for the centers of the 16 [SIII] survey fields
and their dates of observations are given in Table 1.  Exposure times were
40 minutes for the [SIII] on-band, split into two 20-minute exposures
to provide cosmic ray rejection, and 2 minutes for the ``off-band''
RG-850 exposures, also split into two exposures.  The sky was never
reliably photometric during the survey, and so we cannot report fluxes
for the nebulae at [SIII].

All images were reduced in the usual manner with the CCDPROC task
in the IRAF package using twilight sky images for flat-fielding.
The [SIII] and RG-850 images were aligned, and the RG-850 image was
scaled and subtracted from the [SIII] image. The resulting difference
image was visually scanned for positive residuals that signal a [SIII]
emission-line candidate. Coordinates for the candidates were derived
using the IRAF FINDER package, referencing the coordinate system to the HST
Guide Star Catalog. Due to the coarse pixel sampling and the extended
nature of the objects, the derived coordinates have accuracies limited to
$\sim$5\arcsec. The list of candidates is given in Table 2, along with
their diameters, their H$\alpha$+[NII] fluxes when available (see next
section), an estimate of the fraction of flux that is purely H$\alpha$
as derived from the PN spectrum, and their radio fluxes
when available (Van de Steene \& Jacoby 2001). Finding charts 
are presented in the Appendix in Fig. \ref{fig1} - \ref{fig9}; 
the H$\alpha$ images are presented when
available since their resolution is far superior to the [SIII] images.
Diameters in Table 2 are also preferentially taken from the H$\alpha$ images.
We refer to this set of objects with the designation of ``JaSt'' for consistency
with previous papers.

\begin{table}
\caption{Field Centers for the Near-IR PN Survey}
\begin{flushleft}
\begin{tabular}{lccc}
\hline
Field & $\alpha(2000)$ & $\delta(2000)$ & UT Date  \\
\hline
 1  & 17 36 37  & -29 23 27 & 14 July 1994 \\
 2  & 17 38 57  & -28 35 19 & 14 July 1994 \\
 3  & 17 41 14  & -27 47 02 & 14 July 1994 \\
 4  & 17 43 29  & -26 58 35 & 15 July 1994 \\
 5  & 17 40 19  & -29 53 54 & 15 July 1994 \\
 6  & 17 42 37  & -29 05 31 & 15 July 1994 \\
 7  & 17 44 54  & -28 16 59 & 15 July 1994 \\
 8  & 17 47 08  & -27 28 18 & 19 June 1995 \\
 9  & 17 44 03  & -30 23 57 & 19 June 1995 \\
10  & 17 46 20  & -29 35 20 & 19 June 1995 \\
11  & 17 48 35  & -28 46 34 & 19 June 1995 \\
12  & 17 50 49  & -27 57 40 & 20 June 1995 \\
13  & 17 47 49  & -30 53 37 & 20 June 1995 \\
14  & 17 50 05  & -30 04 45 & 20 June 1995 \\
15  & 17 52 19  & -29 15 46 & 20 June 1995 \\
16  & 17 54 31  & -28 26 39 & 21 June 1995 \\ 
\hline
\end{tabular}
\end{flushleft}
\end{table}

\subsection{The H$\alpha$ ``Re-Survey''}

A primary goal of this project is to derive the chemical composition
for PNe in the Galactic bulge from their emission-line ratios. 
Because of the generally heavy extinction to these objects, there
are very few emission lines blueward of H$\alpha$; H$\beta$ 
is extremely weak if it is visible at all. Therefore, we felt that
an alternative method of deriving the extinction was necessary in order
to de-redden the emission-line ratios. The approach we will adopt
is to derive the extinction from the ratio of the radio continuum
emission to the Balmer line emission, using H$\alpha$ as the principal
Balmer line. In the next paper of this series, 
we also demonstrate that one can derive a 
reliable reddening from the ratio of the Paschen P10 line to H$\alpha$.

In addition, the photoionization modeling that is used to derive
the chemical compositions is improved greatly if one has an estimate of the
nebular flux and diameter, as these constrain the central star luminosity
and nebula ionization structure.
The central star luminosity is required to estimate its mass,
and consequently, the age of its progenitor. With ages derived indirectly
from the spectrum and photometric properties of the nebulae,
combined with the chemical composition derived directly from the spectrum,
we can deduce the chemical enrichment history for the stars in
the Galactic bulge.

Consequently, we obtained direct images of the survey fields in
H$\alpha$ in order to measure the fluxes and diameters
of the nebulae using the NOAO CCD Mosaic camera
(Muller \etal 1998) on the KPNO (now, WIYN Observatory)
0.9-m telescope. This facility offers a unique combination of 
wide field (59' x 59'), excellent sensitivity, and good sampling (0.43" pixels).
We observed the Galactic center fields listed in Table 1 
on the nights of 10-13 June 1999.  We used an H$\alpha$+[NII] filter having a
central wavelength of 6569 \AA\  and a FWHM of 80 \AA\ and an
R-band filter to serve as an ``off-band'' filter.
To derive the photometric calibration, each night we observed 4-5
of the following spectrophotometric
standard stars: HZ-44, Wolf 1346, BD+28 4211, Feige 66,
Feige 110, PG 1545+035. Flux measurements were made on the difference images
(H$\alpha$ minus R-band) in order to minimize the degree of
stellar contamination.

Because there are gaps between the CCDs, we observed each of the 16 fields
in a dithered pattern of 5 images per field. Each of the 5 exposures
was 360 seconds, for a total on-band exposure of 30 minutes. Each of the
R-band exposures was 36 seconds, for a total ``off-band'' exposure of 3
minutes. The data were reduced using the IRAF suite of tasks in the MSCRED
package (Valdes 1998), which, after the combination of each dithered
set, yields a single image of approximately 61' x 61'. Thus, these fields are
about 28\% smaller in area on the sky than the Schmidt [SIII] fields, and
some PNe were not recovered at H$\alpha$. Those PNe that were recovered
at H$\alpha$ will have a flux entry in Table 2, while those that were
too faint to be recovered have a flux listed as exactly 0.00.  Those PNe that
fell outside the H$\alpha$ fields are flagged by a blank entry as the flux.
Images of those PNe visible at H$\alpha$ are shown in Fig. \ref{fig1} - \ref{fig9} ; if not visible in H$\alpha$, the chart illustrates the [SIII] image.

\begin{table*}
\caption{Properties of New Galactic Bulge PNe}
\begin{flushleft}
\begin{tabular}{rllclccl}
\hline
JaSt & RA (J2000.0) & DEC (J2000) & Diam$^a$ & Log Flux$^b$ & Flux Ratio$^c$ & 6 cm Flux$^d$ & Comment$^e$ \\
  &   &   & (arcsec) & H$\alpha$+[NII] & H$\alpha$/H$\alpha$+[NII] & (mJy) &  \\
\hline
 1 & 17 34 43.64 &  -29 47 05.03 &   5 & -13.98: & 0.532 & 1.6   & \\
 2 & 17 35 00.96 &  -29 22 15.72 &   5 & -13.22  & 0.883 & 4.3   & \\
 3 & 17 35 22.90 &  -29 22 17.58 &   7 & -13.08  & 0.927 & 12.4  & \\
 4 & 17 35 37.47 &  -29 13 17.67 &   8 & -13.17  & 0.889 & 4.1   & \\
 5 & 17 35 52.51 &  -28 58 27.95 &   4 & -13.11  & 0.850 & 12.7  & \\
 6 & 17 36 38.71 &  -29 17 22.05 & --- &   0.00  &       &       & \\
 7 & 17 38 26.69 &  -28 47 06.48 &  13 & -13.41  & 0.246 &       & \\
 8 & 17 38 27.74 &  -28 52 01.31 &   8 & -13.39  & 0.895 & 3.8   & \\
 9 & 17 38 45.64 &  -29 08 59.27 &   9 & -13.83  & 0.426 & 4.0   & \\
10 & 17 38 47.51 &  -29 01 15.17 &   5 &   0.00  &       &       & \\
11 & 17 39 00.55 &  -30 11 35.23 &  16 & -14.62: & 0.36: & 3.9   & \\
12 & 17 38 59.28 &  -28 46 41.87 &   2 & -14.09: &       &       & \\
13 & 17 39 01.80 &  -29 38 58.07 & --- &         &       &       & \\
14 & 17 39 12.74 &  -29 14 10.74 & --- &         &       &       & \\
15 & 17 39 17.11 &  -29 01 36.24 & --- &         &       &       & \\
16 & 17 39 22.70 &  -29 41 46.08 &   4 & -13.83  & 0.665 & 27.0  & \\
17 & 17 39 31.32 &  -27 27 46.78 &   8 & -12.59  & 0.702 & 10.0  & K6-7 \\
18 & 17 39 34.42 &  -29 08 35.39 & --- &         &       &       & \\
19 & 17 39 39.38 &  -27 47 22.58 &   6 & -12.84  & 0.965 & 6.8   & K6-8 \\
20 & 17 39 50.59 &  -29 04 48.36 &   5 &   0.00  &       &       & \\
21 & 17 39 52.92 &  -27 44 20.54 &  21 & -13.22: & 0.701 &       & \\
22 & 17 40 11.98 &  -30 29 15.65 & --- &         &       &       & \\
23 & 17 40 23.17 &  -27 49 12.04 &   3 & -13.49  & 0.500 & 3.7   & \\
24 & 17 40 28.23 &  -30 13 51.30 & --- &   0.00  & ---   & 16.0  & \\
25 & 17 40 27.83 &  -29 30 06.05 & --- &   0.00  &       &       & \\
26 & 17 40 33.52 &  -29 46 14.98 &  11 &   0.00  & 0.745 & 14.1  & \\
27 & 17 40 42.34 &  -28 12 31.90 &   5 & -13.82  & 1.000 &       & \\
29 & 17 41 01.64 &  -28 58 10.87 & --- &   0.00  &       &       & \\
30 & 17 41 02.27 &  -28 46 26.68 & --- &   0.00  &       &       & \\
31 & 17 41 27.93 &  -28 52 51.61 &   7 & -14.49: & 0.790:& 11.5  & \\
32 & 17 41 43.91 &  -27 33 52.32 &   6 &   0.00  &       &       & \\
33 & 17 41 55.57 &  -29 37 28.89 &   5 &         &       &       & \\
34 & 17 41 54.80 &  -27 03 20.33 &   6 & -13.78  & 0.960 & 1.7   & \\
35 & 17 42 08.71 &  -28 55 37.05 &  28 &         &       &       & \\
36 & 17 42 25.20 &  -27 55 36.36 &   5 & -13.03  & 0.872 & 31.1  & \\
37 & 17 42 28.60 &  -30 09 34.93 &  10 &   0.00  & ---   & 13.5  & IRAS17392-3008 \\
38 & 17 42 32.41 &  -27 33 15.18 &   8 & -13.06: & 0.449 & 2.8   & \\
39 & 17 42 45.14 &  -26 29 20.34 & --- &   0.00  &       &       & \\
40 & 17 42 50.16 &  -28 44 14.69 &   7 &   0.00  &       &       & \\
41 & 17 42 49.96 &  -27 21 19.68 &   7 & -12.73  & 0.906 & 16.7  & \\
42 & 17 43 17.06 &  -26 44 17.67 &   7 & -12.81  & 0.878 & 12.9  & K6-10 \\
43 & 17 43 21.78 &  -28 45 18.68 & --- &   0.00  &       &       & \\
44 & 17 43 23.48 &  -27 34 06.03 &   7 & -13.04  & 0.649 & 3.8:  & \\
45 & 17 43 23.44 &  -27 11 16.91 &  33 & -12.85: & 0.185 &       & \\
46 & 17 43 30.43 &  -26 47 32.33 &   6 & -12.29  & 0.564 & 20.8  & \\
47 & 17 43 33.61 &  -28 18 45.49 &  15 &         &       &       & \\
48 & 17 43 59.26 &  -27 44 32.56 & --- &         &       &       & \\
49 & 17 44 04.34 &  -28 15 57.86 &   6 &         & ---   & 8.2   & \\
50 & 17 44 33.97 &  -26 24 57.10 &  15 &         &       &       & \\
51 & 17 44 40.12 &  -29 22 24.31 & --- &   0.00  &       &       & \\
52 & 17 44 37.30 &  -26 47 25.23 &   5 & -12.47  & 0.665 & 24.0  & \\
53 & 17 44 39.78 &  -26 56 15.66 &   6 &   0.00  &       &       & \\
54 & 17 45 11.06 &  -27 32 36.80 &  15 &   0.00  & ---   & 82.3  & IRAS17420-2731 \\
55 & 17 45 37.36 &  -27 01 18.44 &  10 & -13.18  & 0.485 & 20.0  & \\
56 & 17 45 47.05 &  -27 30 42.03 &   6 &   0.00  & 0.613 & 14.1  & \\
57 & 17 46 30.97 &  -30 16 52.68 & --- &         &       &       & \\
58 & 17 46 52.20 &  -30 37 42.83 &  14 & -14.26: & 0.469 & 29.5  & \\
59 & 17 47 09.66 &  -28 31 16.42 &   9 &   0.00  &       &       & \\
60 & 17 47 53.91 &  -29 36 49.67 &   8 &   0.00  & ---   & 6.5   & IRAS17420-2731 \\
\hline
\end{tabular}
\end{flushleft}
\end{table*}
\addtocounter{table}{-1}
\begin{table*}
\caption{Properties of New Galactic Bulge PNe (continued)}
\begin{flushleft}
\begin{tabular}{rllclccl}
\hline
JaSt & RA (J2000.0) & DEC (J2000) & Diam$^a$ & Log Flux$^b$ & Flux Ratio$^c$ & 6 cm Flux$^d$ & Comment$^e$ \\
  &   &   & (arcsec) & H$\alpha$+[NII] & H$\alpha$/H$\alpha$+[NII] & (mJy) &  \\
\hline
61 & 17 48 23.02 &  -29 40 44.65 &   7 &   0.00  &       &       & \\
62 & 17 48 46.10 &  -27 57 38.61 &   6 &   0.00  &       &       & \\
63 & 17 48 46.27 &  -27 25 37.22 &  10 & -13.37  & 0.261 & 3.9   & \\
64 & 17 48 56.04 &  -31 06 41.95 &   5 & -13.00  & 0.841 & 28.3  & \\
65 & 17 49 20.02 &  -30 36 05.57 &   5 & -12.66  & 0.941 &       & IRAS17461-3035 \\
66 & 17 49 22.15 &  -29 59 27.02 &   5 & -12.86  & 0.902 & 65.3  & \\
67 & 17 49 28.10 &  -29 20 47.56 & --- &         & 1.00: & 27.8  & \\
68 & 17 49 50.87 &  -30 03 10.47 &   5 & -13.96  & 0.953 & 5.9   & \\
69 & 17 50 10.04 &  -29 19 05.14 &  14 & -12.81  & 0.380 & 4.0:  & \\
70 & 17 50 21.07 &  -28 39 02.46 &  11 &   0.00  & ---   & 13.0  & ?IRAS17471-2838
\\
71 & 17 50 23.32 &  -28 33 10.95 &   5 & -13.74  & 0.577 & 34.0  & \\
72 & 17 50 29.06 &  -30 03 41.75 &  22 & -13.33: &       &       & \\
73 & 17 50 47.74 &  -29 53 16.01 &   7 & -12.49  & 0.648 & 13.7  & IRAS17475-2952 \\
74 & 17 50 46.85 &  -28 44 35.42 &   5 & -13.90: & 0.611 & 22.7  & \\
75 & 17 50 48.08 &  -29 24 43.60 &   5 & -13.21  & 0.692 & 22.5  & IRAS17476-292 \\
76 & 17 50 56.47 &  -28 31 24.63 & --- &         & 0.732 & 7.4   & \\
77 & 17 51 11.65 &  -28 56 27.20 &  10 & -12.56  & 0.922 & 49.0  & ?IRAS17480-2855 \\
78 & 17 51 24.68 &  -28 35 40.34 &  11 &         & 0.870 & 12.3  & \\
79 & 17 51 53.63 &  -29 30 53.41 &   5 & -12.54  & 0.545 & 4.1   & \\
80 & 17 51 55.54 &  -27 48 02.46 &  25 & -12.46: & 0.239 &       & \\
81 & 17 52 04.35 &  -27 36 39.28 & --- &   0.00  & 0.135:& 20.8  & IRAS17480-2855 \\
82 & 17 52 05.08 &  -28 05 49.67 &  12 &         &       &       & \\
83 & 17 52 45.17 &  -29 51 05.21 & --- &         & 0.673 & 4.0   & \\
84 & 17 52 46.17 &  -28 48 38.80 &  20 & -13.35: &       &       & \\
85 & 17 52 49.05 &  -29 41 54.92 &  15 & -12.26  & 0.614 & 5.3   & \\
86 & 17 52 52.20 &  -29 30 00.07 &   7 & -12.80  & 0.865 & 8.5   & \\
87 & 17 53 00.28 &  -28 44 15.50 &  15 & -13.45: &       &       & \\
88 & 17 53 00.89 &  -29 05 44.08 &  14 & -12.85: & 0.735 &       & \\
89 & 17 53 06.73 &  -28 18 09.95 &   7 & -13.02  & 0.902 & 8.5   & \\
90 & 17 53 17.77 &  -28 04 33.20 &   5 & -13.93: & 0.777 & 0.97  & \\
91 & 17 53 20.48 &  -29 40 32.30 &  17 &   0.00  &       &       & \\
92 & 17 53 19.81 &  -28 27 14.67 &   5 & -13.93  & 0.523 & ---   & \\
93 & 17 53 24.14 &  -29 49 48.45 &  14 &         & 0.864 & 6.3   & \\
94 & 17 53 33.15 &  -28 35 56.34 & --- &   0.00  &       &       & \\
95 & 17 53 35.38 &  -28 28 51.02 &   9 & -12.30  & 0.570 & 5.5   & \\
96 & 17 53 57.16 &  -29 20 14.97 &  25 & -12.45: &       &       & \\
97 & 17 54 13.36 &  -28 05 16.82 &   7 & -13.25  & 0.677 & 11.5  & \\
98 & 17 55 46.39 &  -27 53 38.91 & --- &         & 0.726 & 21.8  & IRAS17480-2855 \\
\hline
\end{tabular}
\ntd{$^a$ Diameters are taken from the H$\alpha$ images when
possible. Due to poor resolution, the [SIII] diameters are uncertain;
those less than 5\arcsec\ are indicated by ``---".}
\ntd{$^b$ Entries of 0.00 indicate that the PN was not
recovered in the H$\alpha$ imaging.
Blank entries indicate that the PN was not
within the H$\alpha$ imaging survey area. Entries with ``:" have large
uncertainties due to contamination from the residuals of stars, or due to
the faintness of the nebula.}
\ntd{$^c$ Entries of ``---" indicate that the H$\alpha$ line
could not be measured in our spectra. 
Blank entries indicate that no spectrum was
obtained and therefore the PN has not been confirmed spectroscopically.
Entries with ``:" are uncertain due to weak [NII] lines in the spectrum.
Spectra were obtained either at the ESO/La Silla 1.5-m or CTIO 4-m telescopes.}
\ntd{$^d$ Radio fluxes were 
obtained at the Australia Telescope Compact Array
and are reported in Van de Steene \& Jacoby (2001)}
\ntd{$^e$ K-series entries are candidates initially suspected by
Kohoutek (1994). Close coincidences ($<$10\arcsec\ )
with IRAS sources are indicated; 2 IRAS coincidences 
with somewhat larger separations (10-15\arcsec\ ) are shown with ``?".}
\end{flushleft}
\end{table*}

\subsection{New PN Candidates in H$\alpha$}

At modest levels of extinction, H$\alpha$ provides good detection
sensitivity for finding faint PNe. To guard against emission-line stars,
only objects that are non-stellar should be considered viable
PN candidates. With these criteria, 20 additional 
PN candidates were identified
in the H$\alpha$ survey. Until confirmation spectra of these are obtained,
we consider these objects to be candidate PNe.
On the other hand, the morphologies of  at least half of the 
H$\alpha$ candidates look very much like true PNe.

Table 3 lists the new H$\alpha$ candidates, their coordinates, 
diameters, and H$\alpha$+[NII] fluxes. We refer to this new set of
objects with the designation ``JaSt 2'' to distinguish them
from the [SIII] sample. Fig. \ref{fig10} - \ref{fig13} 
provide finding charts.
As Fig. \ref{fig14}  illustrates, the H$\alpha$ candidates are found
predominantly at the most negative latitudes and in selected regions, 
presumably where the extinction is least severe.

\begin{table}
\caption{Optical positions and H$\alpha$ fluxes for PNe Discovered in H$\alpha$ Survey}
\begin{flushleft}
\begin{tabular}{lllcl}
\hline
JaSt 2 & RA  & DEC         & Diam & Log Flux$^a$  \\
 & (J2000.0) & (J2000.0) & (arcsec) & H$\alpha$+[NII] \\
\hline
  1$^b$ & 17 34.29.3 & -29 02 03.0    &  50 & -12.94    \\
  2 & 17 38 44.8 & -28 06 44.4  &  12 & -13.78   \\
  3 & 17 41 09.6 & -28 28 00.0  &   6 & -13.70   \\
  4 & 17 42 27.9 & -27 13 31.4  &  30 & -13.54   \\
  5 & 17 43 48.9 & -26 53 30.0  &  18 & -13.58   \\
  6$^c$ & 17 50 01.5 & -29 33 15.8  &  15 & -12.82    \\
  7 & 17 51 44.0 & -30 12 49.2  &  39 & -12.28   \\
  8 & 17 52 03.5 & -29 16 34.5  &  32 & -12.66   \\
  9 & 17 52 47.6 & -29 17 21.9  &   9 & -13.42   \\
 10 & 17 53 17.2 & -28 36 04.0  &  10 & -12.56   \\
 11 & 17 53 27.0 & -29 08 17.5  &   9 & -13.06   \\
 12 & 17 53 27.5 & -28 11 47.8  &  22 & -13.14   \\
 13 & 17 53 35.6 & -28 52 25.7  &  19 & -13.26   \\
 14 & 17 53 40.7 & -29 42 36.1  &  50 & -12.18   \\
 15 & 17 54 23.5 & -28 34 50.7  &  19 & -12.74   \\
 16 & 17 55 23.0 & -28 20 12.9  &   7 & -13.48   \\
 17 & 17 55 36.5 & -28 25 31.0  &  17 & -12.70   \\
 18 & 17 55 54.9 & -28 05 56.5  &  18 & -13.18   \\
 19 & 17 56 33.8 & -28 30 35.6  &   5 & -13.61   \\
 20 & 17 56 35.9 & -28 57 25.6  &  20 & -12.82   \\ 
\hline
\end{tabular}
\ntd{$^a$ Entries all have very large uncertainties because
of \\ the degree of crowding from stars over the large diameters
\\ of these nebulae.}
\ntd{$^b$ large, possibly bi-polar}
\ntd{$^c$ unusually bright knot just off-center; possibly coincident \\ 
with IRAS 17468-2932 and 1RXS J175001.8-293316 }
\end{flushleft}
\end{table}

\section{Discussion}

\subsection{Verification spectra}

In order to assess the validity of the identification of the candidates
as true PNe, we obtained far red (5000 - 10,000 \AA) spectra of
as many objects as possible during several runs at the ESO 1.5-m during
the June 1995 and July 1996 observing seasons, and a single run at the
CTIO 4-m in June of 1997.  Note that observations at these wavelengths
present some extra challenges due to very severe sky brightness and CCD
fringing. It is difficult, therefore to obtain spectra with accurate sky
subtraction.  The details of these observations are given in the next
paper.

In all, we observed 73 candidates spectroscopically. 
Of these, 63 proved to be genuine
emission-line sources, exhibiting emission at [SIII] $\lambda9532$
as a minimum requirement. The remaining 10 objects that we observed
showed nothing at all, possibly because they were too faint, the coordinates
were inaccurate, the telescope pointing was off, or these are not PNe.
In addition, we did not have enough telescope time to observe
24 candidates.  Based on the very high success rate thus far, though,
most of these are likely to be true PNe as well.
The notes column in Table 2 indicate whether the object was
verified spectroscopically. 

Of the 34 candidates that were not confirmed via spectra, we must be
most skeptical of the 17 candidates that were not recovered either
in the radio or the H$\alpha$ re-surveys (having a 0.00 in column 5
and blank in column 7). Most of these candidates are also
among the faintest, and so, the lack of secondary detections
may be due to the flux limits of the radio and H$\alpha$ surveys.
The exceptions are candidates 53 and 91, for which the [SIII] images
appear moderately bright.

On the other hand, we have very high confidence in those 5 candidates
of the 34 without spectra that were recovered in the H$\alpha$ re-survey
(JaSt 12, 72, 84, 87, 96). These objects all must be emission-line
sources at both [SIII] and H$\alpha$+[NII], plus they have sizes and
morphologies consistent with Galactic center PNe.

For a subset of $\sim12$ objects having relatively
high quality spectra, we see many emission-lines 
that offer insight into the chemical composition
of the nebulae. We will report on these in a subsequent paper.

\subsection{Spatial distribution of the new PNe}

\begin{figure}
\mbox{\epsfxsize=10cm\epsfbox{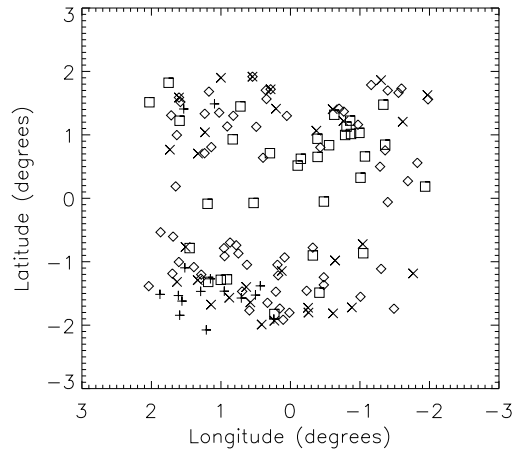}}
\caption{The location of the new PNe discovered in the
[SIII] survey (squares), the new
PNe discovered in the H$\alpha$ ``re-survey'' (plus signs), those
PNe that were confirmed spectroscopically (diamonds) and those PNe
that were previously known in the survey area (crosses). Note the
lack of objects within a half-degree of the Galactic plane, where the
extinction becomes extreme.}
\label{fig14}
\end{figure}

Fig. \ref{fig14}  illustrates the location of the new PNe with respect to their
positions in the Galaxy. The PNe discovered in this survey
extend the identifications of the previously known PNe much closer
to Galactic plane. The central $\pm0.5$ degree 
remains a challenge where the extinction
becomes very large. A few of our new identifications,
though, are right along the plane, suggesting the presence of 
holes in the dust distribution. Alternatively, these objects may
be foreground to the Galactic center.

\subsection{Diameters of the new PNe}

Diameters are only approximate, even for the best cases.
For objects smaller than 5\arcsec\ in [SIII] and 2\arcsec\ in H$\alpha$, 
the spatial
resolution of the images is not adequate for a measurement; we do not present
diameters for these small objects.
Larger objects also have uncertain diameters because the outer
isophotes of PNe can be too faint to see. In addition, PNe are far more
complex than a simple sphere, and so, a single parameter cannot
characterize the size. Nevertheless, even rough estimates
of the nebula sizes are valuable for later photoionization model analysis and for
understanding the selection effects in the identifications.

\begin{figure}
\mbox{\epsfxsize=0.90\columnwidth \epsfbox{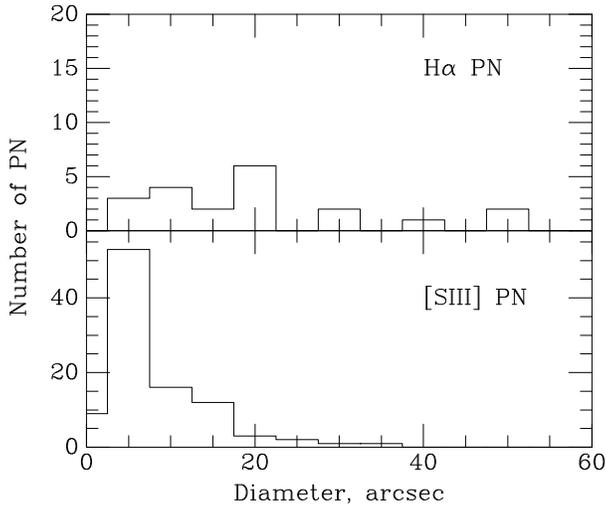}}
\caption{Histograms of the diameter distribution of the new PNe
found in the [SIII] and H$\alpha$ surveys. Some of the objects are as large as
2 pc. Thus, in some cases, we are finding very old PNe. On average, the H$\alpha$
survey is more sensitive to larger PNe.}
\label{fig15}
\end{figure}

Fig. \ref{fig15}  shows the distribution of diameters for the
PNe identified in [SIII] and H$\alpha$. 
Assuming that all these PNe are truly at the distance of the Galactic 
center (7.9 kpc; Eisenhauer \etal 2003), a diameter of
10\arcsec\ corresponds to 0.38 pc. While most of the PNe are fairly small,
a few are quite large, especially those found in H$\alpha$.
In fact, the largest PNe are nearly 2 pc across, demonstrating
that our survey is sufficiently sensitive to find
very old, low surface brightness PNe.

The PNe found only in H$\alpha$ tend to be larger, on average, than those
identified first in [SIII]. This bias suggests that the [SIII] survey is not
nearly as sensitive to the low surface brightness emission. Because the
sky brightness at 9530 \AA\ is much greater than at 6560 \AA, and because
the 0.9-m telescope with the Mosaic Camera is about 10 times more sensitive
than the smaller Schmidt and its thick CCD, the greater depth of the
H$\alpha$ survey is not surprising.

\subsection{OH/IR Stars}
\begin{figure}
\mbox{\epsfxsize=10cm \epsfbox{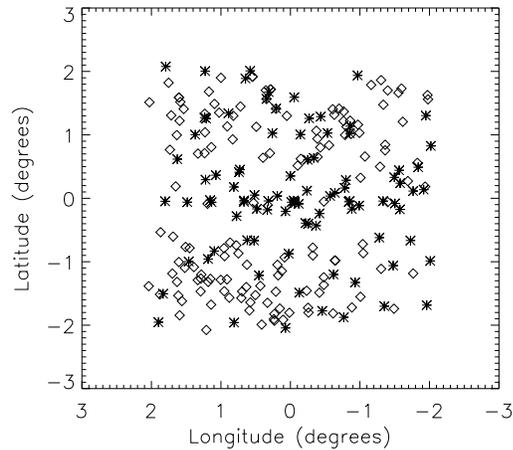}}
\caption{The location of known OH/IR stars (asterisks) compared to 
the new PNe discovered in the [SIII] and H$\alpha$ surveys (diamonds).}
\label{fig16}
\end{figure}

Fig. \ref{fig16}  shows the Galactic distribution of OH/IR stars from Sevenster \etal 
(1997) superposed on our PN distribution. These 1612 Mhz OH maser data 
extend right into the plane because extinction from dust is not a factor. 
Outside the central $\pm0.5$ degree, the
distribution of OH/IR stars is comparable to that of the PNe.
The very high detection rate of OH/IR stars within $\sim$0.25 degrees of
the plane ($\pm30$ pc), though, exceeds any reasonable expectation that all
of these
can be progenitors of PNe. Instead, some of these must be derived from a very
different population. Consequently, 
we caution against the general use of OH/IR stars as tracers of post-AGB stars.

\subsection{The Number of PNe In the Galactic Bulge}

Based on the density of IRAS and OH/IR sources having IR colors similar to PNe,
we anticipated that $\sim320$ PNe are present in our fields within
the Galactic bulge.  The expected number of PNe in our survey region can
also be estimated from the luminosity of the Galactic bulge and
the production rate of PNe as seen in other galaxies.
Dwek \etal (1995) estimated the bolometric luminosity of the central 10 x 10
degrees of the bulge at 5.3 x 10$^9$ L$\sun$. We used their model G0,
corrected for its underestimation relative to the data within 3 degrees
of the center, to
determine that the luminosity fraction appropriate to our 4 x 4 degree survey 
is 19.2\%. This corresponds to 1.02 x 10$^9$ L$\sun$. 
However, Dwek \etal adopted
a distance to the Galactic center of 8.5 kpc, whereas more recent estimates
(e.g., Eisenhauer \etal 2003) find this value to be 7.9 kpc.
Correcting for this distance differential, we adopt a bolometric 
luminosity for our survey region of 0.89 x 10$^9$ L$\sun$. 

For nearby galaxies, Ciardullo (1995) finds that the production rate for bright 
PNe (within 2.5 mags of the most luminous PNe in the galaxy)
varies between $\sim 10$ x $10^{-9}$ and $\sim 50$  x $10^{-9}$ PNe per unit
of bolometric solar luminosities.  The latter number represents the case 
where all stars become PNe and is approached for young populations (e.g.,
LMC and SMC). For the older populations typically seen near the centers
of galaxies, this rate is $\sim 20$ x $10^{-9}$. 
Thus, the Galactic bulge is expected to have $\sim 18$ bright PNe. 
The luminosity function for PNe can be used to extrapolate this number 
to include the faint PNe, but with significant uncertainty. 
That correction factor is 10.1, suggesting a total of $\sim 180$
PNe in the bulge survey zone.  However, this number could be as high
as 450 if the Galactic Center is dominated by a young population of stars.

Because the uncertainties in both estimation methods are difficult
to establish, we simply average these two numbers. Thus, we expect
there to be 250 PNe in the surveyed region of the Galactic bulge.

Our near-IR search identified 140 PNe, for which $\sim$25\% were already
known. The follow-up H$\alpha$ survey identified another 20 candidates, bringing
the total count to 160 PNe, or 64\% of the $\sim250$ PNe expected.
There must certainly be fainter, more extincted, PNe to be found, as evidenced
by our poor detection rate within 0.5 degrees of the Galactic plane.
Nevertheless, we now have identified nearly two-thirds
of the predicted total number of PNe.

\section{Conclusions}

With this near-IR survey for PNe close to the Galactic center, 
we have shown that:

1. Highly extincted PNe can be found effectively by surveying in the
emission line of [SIII] $\lambda$9532, especially if augmented by
a survey in H$\alpha$.

2. The census of PNe within 2 degrees of the Galactic center
is about two-thirds complete.

3. There are now many target PNe for follow-up chemical composition analysis
but most of these are so highly reddened that their blue emission lines
are likely to be unobservable. Thus, optical studies will be
severely hampered except for a few of the brighter, less extincted, objects.
We will present our spectroscopic results for several of these PNe
in a subsequent paper. The remaining PNe to be found in the inner Galactic 
bulge will be even fainter and/or more seriously reddened.

\begin{acknowledgements}

We wish to thank Stuart Pottasch who initially suggested that we carry out
a survey of this nature.

\end{acknowledgements}

\appendix
\section{Finding Charts}

\begin{large}
{\bf The full ps version of this paper including these finding charts
can be found at: http://homepage.oma.be/gsteene/publications.html}
\end{large}

\begin{figure*}
\caption{Finding charts for the new Galactic bulge PN
identified as number 1-12. Each chart is 2 arcmin on a side. Each
PN is denoted by an arrow (2 arrows for
large or ambiguous cases) and is centered in the box, 
except when it was near the very edge of the survey image.
The H$\alpha$ images are shown whenever possible because of their
superior spatial resolution. The [SIII] images are used for PN
6, 9, and 10.}
\label{fig1}
\end{figure*}

\begin{figure*}
\caption{As for Fig. \ref{fig1}, but showing the finding
charts for PN 13-24. [SIII] images are shown for PN 13, 14, 15, 18,
20, 22, and 24.}
\label{fig2}
\end{figure*}

\begin{figure*}
\caption{As for Fig. \ref{fig1}, but showing the finding
charts for PN 25-36. [SIII] images are shown for PN 25, 26, 29, 30,
32, 33, and 35. Note that PN 28 is not shown, as it was later
removed from the list of confirmed PN.}
\label{fig3}
\end{figure*}

\begin{figure*}
\caption{As for Fig. \ref{fig1}, but showing the finding
charts for PN 37-48. [SIII] images are shown for PN 37, 39, 40, 43,
47, and 48.}
\label{fig4}
\end{figure*}

\begin{figure*}
\caption{As for Fig. \ref{fig1}, but showing the finding
charts for PN 49-60. [SIII] images are shown for 
PN 49, 50, 51, 53, 54, 56, 57, 59, and 60.}
\label{fig6}
\end{figure*}

\begin{figure*}
\caption{As for Fig. \ref{fig1}, but showing the finding
charts for PN 61-72. [SIII] images are shown for 
PN 61, 62, 67, and 70.}
\label{fig6}
\end{figure*}

\begin{figure*}
\caption{As for Fig. \ref{fig1}, but showing the finding
charts for PN 73-84. [SIII] images are shown for 
PN 76, 77, 78, 81, 82, and 83.}
\label{fig7}
\end{figure*}

\begin{figure*}
\caption{As for Fig. \ref{fig1}, but showing the finding
charts for PN 85-96. [SIII] images are shown for 
PN 91, 93, and 94.}
\label{fig8}
\end{figure*}

\begin{figure*}
\caption{As for Fig. \ref{fig1}, but showing the finding
charts for PN 97-98. [SIII] images are shown for 
PN 98.}
\label{fig9}
\end{figure*}

\begin{figure*}
\caption{Finding charts for the PN candidates discovered
in the follow-up H$\alpha$ survey identified as number 1-12. Otherwise,
these are similar to Fig. \ref{fig1}.}
\label{fig10}
\end{figure*}

\begin{figure*}
\caption{As for Fig. \ref{fig10}, but for PN 13-20.}
\label{fig11}
\end{figure*}

\begin{figure*}
\caption{As for Fig. \ref{fig10}, again showing candidates 1-12,
but now with the continuum image subtracted to illustrate better 
these very low surface brightness PNe.}
\label{fig12}
\end{figure*}

\begin{figure*}
\caption{As for Fig. \ref{fig12}, but for PN 13-20.}
\label{fig13}
\end{figure*}

\end{document}